\documentclass[twocolumn,english,pra,amssymb,aps,superscriptaddress,showpacs,twocolumn,amsmath,showkeys,floatfix,]{revtex4-1}
\usepackage[T1]{fontenc}
\usepackage[latin9]{inputenc}
\usepackage{geometry}
\usepackage[active]{srcltx}
\usepackage{amsmath}
\usepackage{graphicx}
\usepackage{esint}
\usepackage{epsfig}
\usepackage{epstopdf}
\usepackage{physics}
\usepackage{color}
\usepackage{natbib}
\usepackage{upgreek}
\usepackage{mathtools}
\usepackage{soul}
\usepackage[dvipsnames]{xcolor}
\usepackage{ulem}
\usepackage{color}
\definecolor{blue}{rgb}{0,0,1}
\definecolor{green}{rgb}{0,1,0}
\definecolor{purple}{rgb}{0.5,0,1}
\makeatletter
\usepackage{babel}

\makeatother

\usepackage{babel}
\begin{document}

\title{Birefringence and dichroism effects in the spin noise spectra of a spin-1 system}


\author{S. Liu}
\affiliation{Universit\'e Paris-Saclay, CNRS, Ecole Normale Sup\'erieure Paris-Saclay, CentraleSup\'elec, LuMIn, Orsay, France}
\affiliation{East China Normal University, State Key Laboratory of Precision Spectroscopy, Shanghai, China}
\author{P. Neveu}
\affiliation{Universit\'e Paris-Saclay, CNRS, Ecole Normale Sup\'erieure Paris-Saclay, CentraleSup\'elec, LuMIn, Orsay, France}
\author{J. Delpy}
\affiliation{Universit\'e Paris-Saclay, CNRS, Ecole Normale Sup\'erieure Paris-Saclay, CentraleSup\'elec, LuMIn, Orsay, France}
\author{L. Hemmen}
\affiliation{Universit\'e Paris-Saclay, CNRS, Ecole Normale Sup\'erieure Paris-Saclay, CentraleSup\'elec, LuMIn, Orsay, France}
\author{E. Brion}
\affiliation{CNRS, Universit\'e de Toulouse III Paul Sabatier, LCAR,  IRSAMC, Toulouse, France}
\author{E. Wu}
\affiliation{East China Normal University, State Key Laboratory of Precision Spectroscopy, Shanghai, China}
\author{F. Bretenaker}
\affiliation{Universit\'e Paris-Saclay, CNRS, Ecole Normale Sup\'erieure Paris-Saclay, CentraleSup\'elec, LuMIn, Orsay, France}
\author{F. Goldfarb}
\affiliation{Universit\'e Paris-Saclay, CNRS, Ecole Normale Sup\'erieure Paris-Saclay, CentraleSup\'elec, LuMIn, Orsay, France}
\affiliation{Institut Universitaire de France (IUF)}

\begin{abstract}
{We perform spin noise spectroscopy experiments in  metastable helium atoms at room temperature, with a probe light whose frequency is blue detuned from the $D_0$ line. Both circular birefringence fluctuations (Faraday noise) and linear birefringence fluctuations (ellipticity noise) are explored theoretically and experimentally. In particular, it is shown that in both cases but for different optical detunings, two noise resonances are isolated at the Larmor frequency and at twice the Larmor frequency with a behaviour, which strongly depends on the orientation of the probe field polarization. The simple structure of metastable helium allows us to probe, model and explain the changes in the behavior of these peaks in terms of circular and linear dichroisms and birefringences as well as in terms of spin oscillation modes.
}

\end{abstract}

\maketitle

\section{Introduction}
\begin{figure}
    \centering
    \includegraphics[width=\columnwidth]{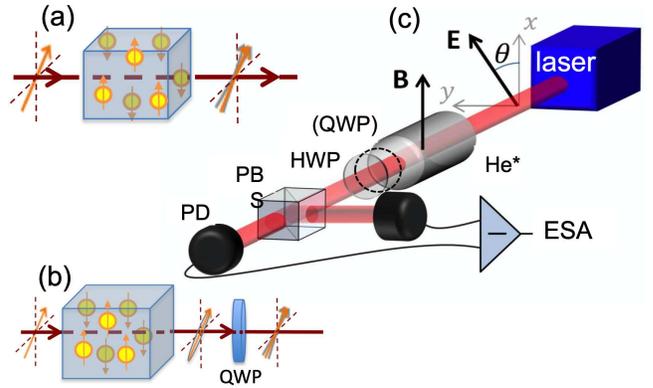}
    \caption{(a) In a paramagnetic ensemble, random spin fluctuations induce a random Faraday rotation of a linearly polarized probe light, which can be converted into an intensity noise using a polarizing beamsplitter (PBS) followed by a balanced detection (RND set-up). (b) A properly oriented quarter wave plate (QWP) can be added to convert an ellipticity noise into an orientation noise (END set-up). (c) Experimental set-up: a half-wave plate (HWP) allows to have an equal average intensity on both photodiodes (PD). The zero frequency centered noise resonance is shifted by a transverse DC magnetic field so that the noise peak is detected around the Larmor frequency $\omega_{\mathrm{L}}$. The QWP can be added or removed, to get the END or RND set-up respectively.} 
    \label{fig.ManipSNS}
\end{figure}

In magnetic systems, the spectroscopy of fundamental noise due to random spin fluctuations, called Spin Noise Spectroscopy (SNS), can be optically performed by measuring the associated fluctuations of the Faraday-like rotation experienced by a linearly polarized probe beam propagating through the sample \cite{Sinitsyn16} (see Fig. \ref{fig.ManipSNS}(a)). Although a first experimental effort was initially reported in the early 1980s  \cite{Alexandrov81}, this method has seen a renewed interest in the past 20 years, due to advances in narrow linewidth lasers and developments in low noise electronics required for spectrum analysis \cite{Crooker04}: second order (and higher order) correlators of atomic spin fluctuations were shown to give access to classical (and purely quantum) signatures of the interaction of the system with its environment, and SNS was then used to probe different properties of various media such as thermal atomic vapours, semiconductors or quantum wells \cite{Zapasskii13, Mihaila06, Smirnov17}. SNS is also proposed as a tool for precision magnetometry \cite{Swar18}, using squeezed-light \cite{Lucivero16} or homodyne detection amplification \cite{Sterin18, Petrov18b} to improve its sensitivity. Recent works extend the SNS technique to the study of spatiotemporal correlations by spatial phase-modulation of the optical probe \cite{Cronenberger19} and the measurement of correlations beyond the second order raises a lot of interest, as they can be used to probe the limits of the linear response and fluctuation-dissipation theory and thus give access to new phenomena \cite{Li13}. 

In usual spin noise experiments, the noise in the polarization orientation of the probe laser  is frequency shifted by a transverse DC magnetic field. The spin noise then lies in the frequency domain around the Larmor frequency $\omega_{\mathrm{L}}$. It was recently shown that for spins equal or larger than 1, it is also possible to observe some polarization ellipticity noise at frequency $2\omega_{\mathrm{L}}$. This noise  can be explained  phenomenologically as the contribution of spin alignment fluctuations to the noise of the symmetric part of the dielectric susceptibility \cite{Bretenaker2019,Fomin20}. Fomin\& al. \cite{Fomin20} provided an analytical equation for the polarization dependence behavior of this noise power, which explains qualitatively well the results obtained when the ellipticity noise is recorded far enough from the transition resonance. Nevertheless, some discrepancy remains that was attributed to pumping effects. Moreover, one can wonder what happens when the probe detuning is reduced.  We thus propose here to explore experimentally and theoretically the changes in both the orientation and ellipticity spin noise spectra when the laser beam is set closer to the atomic transition. To this aim, we choose to use metastable helium atoms, which have the advantage of not exhibiting any hyperfine structure and thus to exhibit a very simple level scheme. The goal of the present paper is thus to explore the properties of the SNS in this relatively pure and simple closed $J=1\rightarrow J=0$ atomic transition, in order to put into evidence the peculiarities of spin noise in the case of a spin 1 system like the $\ket{2^3S_1}$ state. 

A first part gives details on the experimental set-up that allows to record both types of spin noise spectra, before showing results far from resonance, where optical pumping effects can be neglected. We then present the numerical model that we used and compare theoretical and experimental results when the noise origin is limited to the transit noise. In the last part, the same model is applied to explain experimental results obtained closer to resonance. Guided by this model we then provide physical interpretations of the results in terms of existence of several spin oscillation modes, which lead to linear or circular birefringence and dichroism.

\section{Experimental set-up}
Contrary to alkali atoms, the $\ket{2^3S_1}\leftrightarrow \ket{2^3P}$ transition of metastable helium has a fine structure only, with three absorption lines separated by a frequency difference larger than the 0.8\,GHz room temperature Doppler broadening \cite{Goldfarb08}. The highest excited level is the $\ket{2^3P_0}$ state, which is nearly 30\,GHz away from the intermediate $\ket{2^3P_1}$ one: one can then choose to blue detune the frequency of a probe beam from the corresponding $D_0$ line in order to probe this transition in a purely dispersive and perturbative manner without being significantly disturbed by the influence of the two other lines. 

Figure \ref{fig.ManipSNS}(c) shows a schematics of the experimental setup. A linearly polarized laser beam at $1.083\,\upmu$m (CYFL-kilo ultra low noise keopsys fiber laser), corresponding to the $\ket{2^3S_1}\leftrightarrow \ket{2^3P}$ optical transitions, with a diameter  about 0.6\,mm, is sent into a cell filled with 1\,Torr of $^4$He at room temperature. A radiofrequency discharge at 27\,MHz generates a plasma in the cell, so that a fraction of helium atoms are excited to the $\ket{2^3S_1}$ metastable state through collisions with electrons, leading to a metastable helium density of about $2\times10^{11}\,\mathrm{cm}^{-3}$. The remaining He atoms then play the role of a buffer gas, which redistributes optical pumping all over the Doppler broadening through collisions \cite{Goldfarb08}. While propagating inside the helium cell, the laser polarization experiences some random Faraday rotation induced by random fluctuations of the populations of the three Zeeman ground state sublevels. A transverse DC magnetic field is added to the atoms in order to shift the spin noise resonance around the Larmor frequency $\omega_{\mathrm{L}}$ \cite{Zapasskii13, Sinitsyn16}. This B-field is created by two rectangular coils, whose dimensions are much larger than the cell, so that it can be considered as homogeneous inside the sample. The relative angle between this magnetic field and the incident polarization direction of the probe beam is denoted as $\theta$.

The $\ket{2^3S_1}$ ground state has three Zeeman sublevels, with a $\pm2.8$\,MHz/Gauss Zeeman shift for the $m=\pm1$ sublevels, respectively. In our experimental conditions, the Larmor frequency lies in the MHz range.

After the cell, a half-wave plate (HWP) followed by a polarizing beamsplitter (PBS) allows to get an equal average intensity on the two photodiodes of a balanced detection, so that their signal difference is zero on average. The associated noise power spectral density can then be recorded using an electrical spectrum analyzer (ESA). To describe the balanced detection scheme, the output field complex amplitude can be written as the sum of a mean value $\mathbf{E}$ and a fluctuating part $\mathbf{e}\left(t\right)$:
\begin{equation}
\mathbf{E}_{\mathrm{out}}\left(t\right)=\mathbf{E}+\mathbf{e}\left(t\right)\,,\label{Eq01}
\end{equation}
where $\mathbf{E}$ is assumed to be real, and the vectorial fluctuations $\mathbf{e}\left(t\right)$ can consist in amplitude, phase or polarization variations, which carry the spin noise information.

\begin{figure}
    \centering
    \includegraphics[width=\columnwidth]{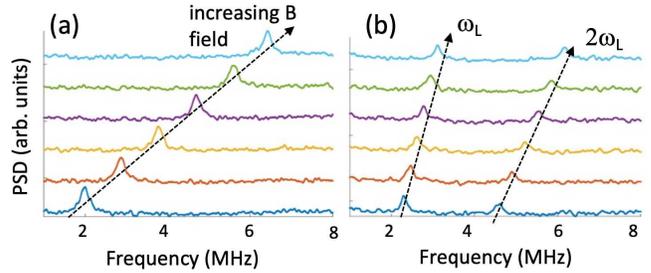}
    \caption{(a) Experimental data recorded with an electrical spectrum analyzer (ESA) for different magnetic fields in the case in the case of the rotation noise detection (RND) set-up, for an angle $\theta=0^{\circ}$ between the input linear polarization and the magnetic field: one resonance is visible at the Larmor frequency $\omega_{\mathrm{L}}$. (b) Experimental data recorded in the case of the ellipticity noise detection (END) set-up, for $\theta\simeq 55^{\circ}$: two resonances are then visible, at $\omega_{\mathrm{L}}$ and $2\omega_{\mathrm{L}}$. Arbitrary offsets are added for a better visibility.}
    \label{fig.1And2Larmor}
\end{figure}

After the HWP, the field sent to the balanced detection has an average polarization at $45^{\circ}$ of the polarizing beamsplitter axes $\mathbf{x}$ and $\mathbf{y}$ :
\begin{align}
\mathbf{E}_{\mathrm{out}}\left(t\right)=\frac{1}{\sqrt{2}}&\left[\left(E+\mathbf{e}_{\|}\left(t\right)+\mathbf{e}_{\perp}\left(t\right)\right)\mathbf{x}\right. \nonumber\\
&+\left.\left(E+\mathbf{e}_{\|}\left(t\right)-\mathbf{e}_{\perp}\left(t\right)\right)\mathbf{y}\right]\,.\label{eq.PerfectLinearField}
\end{align}
where $\mathbf{e}_{\|}\left(t\right)$ and $\mathbf{e}_{\perp}\left(t\right)$ are the components of $\mathbf{e}\left(t\right)$, which are parallel and perpendicular to $\mathbf{E}$, respectively, and $E$ is the amplitude of $\mathbf{E}$. The  signal at the output of the balanced detection is thus proportional to:
\begin{align}
D= &\left|\mathbf{E}_{\mathrm{out}}\left(t\right)\cdot\mathbf{x}\right|^{2}-\left|\mathbf{E}_{\mathrm{out}}\left(t\right)\cdot\mathbf{y}\right|^{2} \nonumber\\
= & 2E\,\mathrm{Re}\left(\mathbf{e}_{\perp}\left(t\right)\right)+{\cal O}\left(\mathbf{e}\left(t\right)^{2}\right)\,.\label{eq.PerfectSignal}
\end{align}

The ESA thus records the power spectral density (PSD) of the field fluctuations polarized orthogonally to the mean output field, and in phase with it. This corresponds to fluctuations of the orientation of a linearly polarized field, which we can call \textit{Faraday noise}. This first order calculation remains valid in the case of a mean field with a small ellipticity. Figure \ref{fig.1And2Larmor}(a) shows examples of experimental noise spectra obtained for increasing transverse magnetic field values, in the case of a 1.5\,mW probe laser blue detuned by 1.5\,GHz: the spin noise resonance frequency, which is centered on the corresponding Larmor frequency, increases with the magnetic field value. 

It is also possible to monitor ellipticity fluctuations by inserting a quarter-wave plate (QWP) just after the cell, as shown in Fig.\,\ref{fig.ManipSNS}(b). While Fomin \& al \cite{Fomin20} used a circularly polarized input beam, we choose to orient the QWP to transform an input elliptic polarization into a linear one: the ellipticity noise is then changed into rotation noise, which can be detected with the same set-up as before. However, the QWP induces a $\pi/2$ phase shift on the fluctuations along the $\mathbf{y}$ axis so that, when the average ellipticity is small and can be neglected, Eq.\,(\ref{eq.PerfectLinearField}) must be replaced by:
\begin{align}
\mathbf{E}_{\mathrm{out}}\left(t\right)=\frac{1}{\sqrt{2}}&\left[\left(E+\mathbf{e}_{\|}\left(t\right)+i\mathbf{e}_{\perp}\left(t\right)\right)\mathbf{x}\right. \nonumber\\
&+\left.\left(E+\mathbf{e}_{\|}\left(t\right)-i\mathbf{e}_{\perp}\left(t\right)\right)\mathbf{y}\right]\,.
\end{align}
Consequently, the signal analyzed by the ESA becomes:
\begin{equation}
D= 2E\;\mathrm{Im}\left(\mathbf{e}_{\perp}\left(t\right)\right)+{\cal O}\left(\mathbf{e}\left(t\right)^{2}\right)\,.
\end{equation}

Instead of the field fluctuations that are in phase with the mean field, this setup thus probes the part of the fluctuations in quadrature with it, i.e., the \textit{ellipticity noise}. In the following, we call this setup the ellipticity noise detection (END) setup, while the first one is called the rotation noise detection (RND) setup. 
These two setups correspond to the two limiting cases explored in \cite{Petrov18b}, where the relative phase between the local oscillator and the signal can be continuously tuned: the balanced detection plays the role of a homodyne detector in which the local oscillator is provided by the mean  field $E$ and the role of the signal is played by the fluctuations $\mathbf{e}_{\perp}$. One can choose between the two signal quadratures by adding or not the QWP as shown in Fig.\,\ref{fig.ManipSNS}(c). It should be noted that we always check that the wave plate is oriented so that the average output polarization is linear and the detection properly balanced. Figure \ref{fig.1And2Larmor}(b) shows examples of experimental noise spectra obtained with this END set-up for increasing transverse magnetic field values with $\theta=55^{\circ}$, in the case of a 1.5\,mW probe laser blue detuned by  1.5\,GHz : there are now two spin noise resonances, at the Larmor frequency $\omega_{\mathrm{L}}$ and at $2\omega_{\mathrm{L}}$. 

\section{Far-detuned experimental results}
\begin{figure}[h]
    \centering
    \includegraphics[width=\columnwidth]{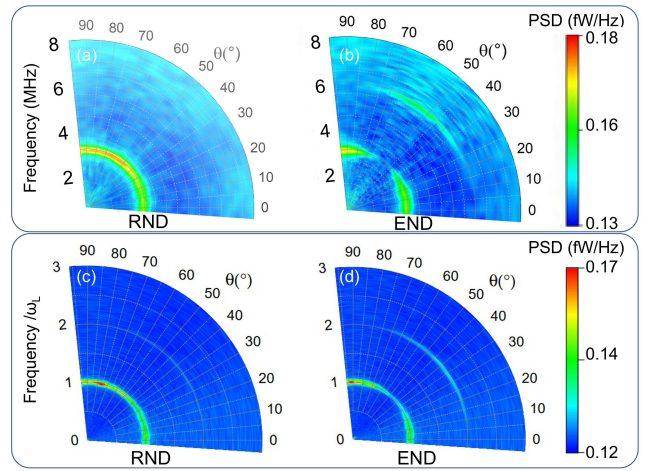}
    \caption{(a,b) Experimental and (c,d) simulated  power spectral densities (PSD) obtained 1.5\,GHz above the $D_0$ transition for a 1.5\,mW laser input power. They are plotted as a function of the input polarization angle $\theta$ from the transverse magnetic field for (a,c) the RND set-up and (b,d) the END set-up. Experimental spectra are recorded every $4^{\circ}$. The ESA resolution and video bandwidths are 91\,kHz and 15\,Hz, respectively. The simulations are  obtained with a spontaneous emission decay rate $\Gamma_0/2\pi=1.6$\,MHz, an optical coherence  decay rate $\Gamma/2\pi=0.8$\,GHz that takes Doppler broadening into account, and equal transit and Raman decay rates $\gamma_{\mathrm{t}}/2\pi=\gamma_{\mathrm{R}}/2\pi=30$\,kHz. The probe beam Rabi frequency is $\Omega_{\mathrm{P}}/2\pi=40$\,MHz} 
    \label{fig.PolarizationDependence}
\end{figure}

The results presented in Fig.\,\ref{fig.1And2Larmor} were obtained for a specific input polarization angles with respect to the magnetic field orientation. More information is gained by scanning  the polarization angle from $\theta=-4^{\circ}$ to $\theta=94^{\circ}$ for both the RND and the END setups. Such SNS scans, with their spectra coded in false colors, are shown in  Fig.\,\ref{fig.PolarizationDependence}(a,b).  They are obtained with a 1.5\,mW probe beam, which is blue detuned far from the resonance, by about 1.5\,GHz from the center of the $D_0$ line, for the (a) RND  and (b) END setups. In the former case, only the peak at  $\omega_L$ is present, and its power does not exhibit any polarization dependence. On the contrary, both resonances can be detected in the case of the END setup, but not for the same input polarization orientations: the resonance at $\omega_\mathrm{L}$ is visible around $\theta\simeq0^{\circ}$ and $\theta\simeq90^{\circ}$, while the peak at $2\omega_\mathrm{L}$ exists for $\theta\simeq50^{\circ}$, when the peak at $\omega_\mathrm{L}$ disappears. 

To understand these observations, and in particular the polarization dependence of the different noise components, we developed a numerical model that is presented in the next section. It will also be used to predict what should happen when the laser is set closer to resonance, as experimentally observed in Section \ref{Sec05}. Finally, a physical explanation of all these different features will be given in Section \ref{Sec06}, at the end of the paper.

\section{Simple numerical model}
We consider the density matrix $\rho$ of the system composed of the three Zeeman sublevels of the $\ket{2^3S_1}$ state and the excited $\ket{2^3P_0}$ level. Its evolution can be modeled by the usual Hamiltonian and dissipative parts, to which we add a fluctuation term $f(t)$ that can be adapted to the physical origin of the noise:
\begin{equation}
    \frac{d\rho}{dt}=\frac{1}{i\hbar}[H,\rho]+\frac{1}{i\hbar}{D}(\rho)+f(t)\ .\label{Eq06}
\end{equation}
Here $H$ is the Hamiltonian and $D(\rho)$ holds for the system relaxations. The Hamiltonian $H$ takes into account the effect of the transverse DC magnetic field through the corresponding Larmor frequency $\omega_{\mathrm{L}}$, and the interaction with the probe light, which couples the $\sigma_\pm$ transitions with Rabi frequencies $\Omega^{\pm}$ and is optically detuned from the optical transition by a frequency $\Delta$:
\begin{equation}
H=\hbar\left(\begin{array}{cccc}
0 & \frac{\omega_\mathrm{L}}{\sqrt{2}} & 0 & \frac{\Omega^{+*}}{\sqrt{3}}\\
\frac{\omega_\mathrm{L}}{\sqrt{2}} & 0 & \frac{\omega_\mathrm{L}}{\sqrt{2}} & 0\\
0 & \frac{\omega_\mathrm{L}}{\sqrt{2}} & 0 & -\frac{\Omega^{-*}}{\sqrt{3}}\\
\frac{\Omega^{+}}{\sqrt{3}} & 0 & -\frac{\Omega^{-}}{\sqrt{3}} & \Delta
\end{array}\right)\\ .\label{Eq07}
\end{equation}
The matrix representation of $H$ in Eq.\,(\ref{Eq07}) is written in the basis $\{\ket{-1}_z, \ket{0}_z, \ket{1}_z,\ket{\mathrm{e}}\}$, where $\ket{-1}_z$, $\ket{0}_z$, $\ket{1}_z$ are the Zeeman sublevels of the lower state of the transition with the quantization axis chosen along $z$, the light propagation axis, and where $\ket{\mathrm{e}}$ is the excited level.

The relaxation term $D(\rho)$ contains the upper level population decay rate $\Gamma_0=2\pi\times1.6\times10^6\,\mathrm{rad.s}^{-1}$, the optical coherence decay rate $\Gamma$, the transit rate $\gamma_{\mathrm{t}}$ and the Raman coherence decay rate $\gamma_{\mathrm{R}}$ of the lower levels. In our experimental conditions, $\gamma_{\mathrm{t}}/2\pi=\gamma_{\mathrm{R}}/2\pi\simeq30$\,kHz. As already shown in many circumstances, the velocity changing collisions in this system are very efficient to redistribute the atoms among the different velocity classes \cite{Figueroa06,Goldfarb08}, leading to the fact that $\Gamma$ can be replaced by the Doppler width in the simulations, i. e. $\Gamma/2\pi=0.8\,\mathrm{GHz}$.

The last term $f(t)$ in Eq.\,(\ref{Eq06}) is introduced to model the source of fluctuations. In principle, all relaxation mechanisms are random by nature and should be taken into account in this term. However, here, for the sake of simplicity, we suppose that the fluctuations arising  from the motion of the atoms in and out of the laser beam dominate, and we introduce them in $f(t)$ in a heuristic manner. Since the lifetime $1/\Gamma_0$ of the upper level is much shorter than the typical transit time $1/\gamma_{\mathrm{t}}$, we neglect the random fluctuations associated with the upper level and write
\begin{equation}
f(t)=\left(\begin{array}{cccc} f_{-1} & k^*_{-1,0} & k^*_{-1,1}& 0 \\  k_{-1,0} & f_{0} & k^*_{0,1} &0 \\  k_{-1,1} & k_{0,1} & f_{1} & 0\\ 0& 0& 0& 0\end{array}\right)\ .\label{Eq08}
\end{equation}

The real diagonal terms $f_j$ with $j=-1,0,1$ correspond to the fluctuations of the populations of the three ground state sublevels. Since the average number of atoms $\overline{N}$ in the interaction region is large, the Poissonian laws governing the fluctuations $f_j\mathrm{d}t$ of the fraction of atoms in state $j$ leaving and entering the interaction volume over a simulation time step $\mathrm{d}t$ are approximated by  Gaussian random variables with a sum of their variances equal to $\sigma^2=2\gamma_{\mathrm{t}}\mathrm{d}t/3\overline{N}$.

The non diagonal terms $k_{i,j}$ with $i,j=-1,0,1$ in Eq.\,(\ref{Eq08}) are complex random variables and correspond to the fluctuations of the Zeeman coherences in the lower level. Since their statistical properties have no reason to depend on the chosen quantization axis, it is quite easy to show that their real and imaginary parts are independant Gaussian variables with equal variances $3\sigma^2/4$ for $\mathrm{Re}(k_{i,j})\mathrm{d}t$ and $\mathrm{Im}(k_{i,j})\mathrm{d}t$: see the supplemental for more information about this point and how to model the polarization parameters.

The results of these simulations are given in Fig.\,\ref{fig.PolarizationDependence} (c,d), in the case of both setups. They reproduce very well the experimental results: it is even visible that similarly to the experimental data, the noise component at $\omega_{\mathrm{L}}$ is stronger close to $\theta=0^{\circ}$ than to $\theta=90^{\circ}$. Nevertheless, one can notice a small discrepancy by about 5 to $10^{\circ}$ between the angular positions of the peaks,  and it is visible that the experimental resonances are  broader than the theoretical ones. The first discrepancy can be explained by the fact that finding the angle $\theta=0^{\circ}$ is not easy and by some possible misorientation of the DC magnetic field. The fact that the experimental noise peaks are broader than the simulated ones can arise from the fact that the Rabi frequency can be underestimated. Indeed, the simulations are based on its average value across the Gaussian beam section (top-hat beam approximation), while in the experiment it is larger at the center of the beam, which can broaden the SNS peaks. Possible magnetic field inhomogeneities can also contribute to this discrepancy.

Fomin \& al. \cite{Fomin20,Fomin21} have already mentioned that two noise resonances can be recorded when the ellipticity noise is probed. Nevertheless, they observed a polarization dependence of the noise peaks that significantly differs from the one observed here (see Fig.\,\ref{fig.PolarizationDependence}) and from their own predictions. This was attributed to the fact that these experiments were performed close to the center of the D2 line of Cs atoms. The associated absorption then induces an ellipticity contribution, as already reported in semiconductors \cite{Poltavtsev14, Yang14}. In the following section, we thus explore the behavior of our system when the probe frequency gets closer to resonance, and the influence of the associated optical pumping processes. 

\section{Spin noise near resonance}\label{Sec05}
\begin{figure}[h]
    \centering
    \includegraphics[width=\columnwidth ]{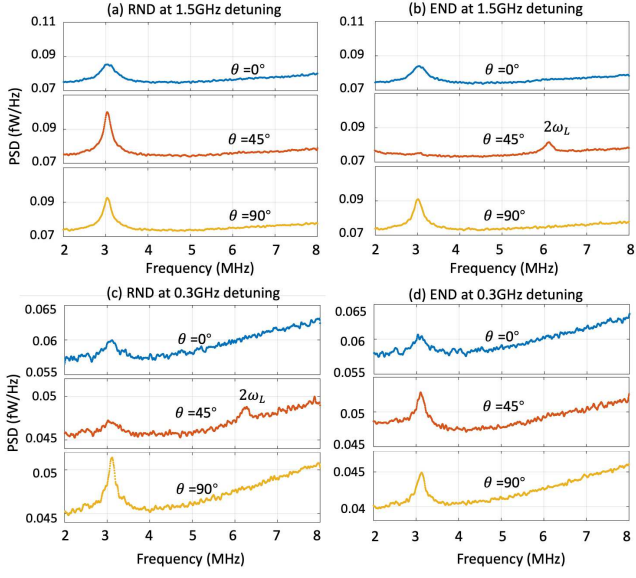}
    \caption{Experimental spectra obtained for $\theta=0^{\circ}$, $45^{\circ}$ and $90^{\circ}$ in the case of (a) the RND set-up and an optical detuning $\Delta=1.5$\,GHz; (b) the END set-up and $\Delta=1.5$\,GHz; (c) the RND set-up and $\Delta=0.3$\,GHz (d) the END set-up and $\Delta=0.3$\,GHz. The laser input power is about 1\,mW.}
    \label{3deg}
\end{figure}

Generally, SNS is performed far away from resonance where the absorption is negligible, in order to avoid to disturb the system. But remaining far from resonance can be difficult in systems exhibiting large inhomogeneous broadening, and some groups already investigated the effect of small optical detunings \cite{Mitsui00,Petrov18a}. In such conditions, absorption cannot be ignored and birefringence effects become negligible compared to dichroism ones, so that the behavior of the SNS resonances changes completely.

Figure \ref{3deg} shows spectra obtained for three different polarization directions $\theta=0^{\circ},45^{\circ},90^{\circ}$, in both RND and END set-up cases. The ones recorded at 1.5\,GHz detuning (see Figs.\,\ref{3deg}(a,b)) are consistent with the theory developed by Fomin \& al \cite{Fomin20}. But when the detuning decreases down to 0.3\,GHz, a peak at  2$\omega_{\mathrm{L}}$ becomes visible in the case of the RND set-up data (see the middle spectrum in Fig.\,\ref{3deg}(c)) while it disappears from the END spectra (see Fig.\,\ref{3deg}(d)). It is also worth noticing that the signal becomes much smaller when the detuning is reduced, because of the absorption that decreases the power reaching the detector. This explains why the detector background noise, which increases with the frequency, becomes much more visible in  Figs.\,\ref{3deg}(c,d) compared with Figs.\,\ref{3deg}(a,b).

\begin{figure}
    \centering
    \includegraphics[width=\columnwidth ]{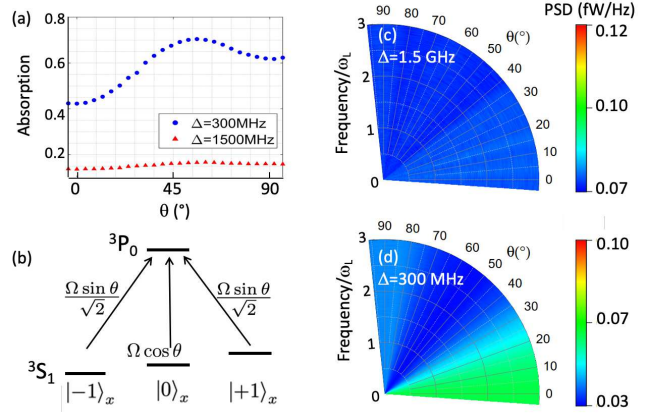}
    \caption{(a) Experimental measurements of the absorption of a 1\,mW probe for a detuning of 1.5\,GHz (red triangles) and 300\,MHz (blue circles). (b) Sub-transitions of the $^3 S_1\rightarrow ~ ^3 P_0$ line, and their excitations by an electric field, which has an angle $\theta$ with respect to the quantization axis and a corresponding total Rabi frequency $\Omega$. The background noise is calculated using these experimental data for optical detunings of (c) 1.5\,GHz and (d) 300\,MHz.}
    \label{ShotNoise}
\end{figure}

It can also be noticed by comparing for example the different spectra in Fig.\,\ref{3deg}(d) that the noise floor at $\theta=0^{\circ}$ is larger than at $\theta=45^{\circ}$ and $90^{\circ}$. This is due to the fact that the detected power depends on $\theta$, because the transverse magnetic field $\mathbf{B}$ makes the vapor transmission depend on the incident polarization direction. Indeed, Fig.\,\ref{ShotNoise}(a) reproduces measurements of the evolution of the cell absorption versus $\theta$, clearly evidencing this effect. The result of this dependence of the detected average power on $\theta$ is a polarization dependence of the shot noise level, as clearly visible in Figs.\,\ref{3deg}(c,d). The dependence of the cell absorption on $\theta$ can be easily understood if one uses a quantization axis along $\mathbf{B}$: Figure \ref{ShotNoise}(b) shows the transitions induced by a light beam linearly polarized at an angle $\theta$ with respect to the quantization axis $x$, from the sublevels $\ket{-1}, \ket{0}, \ket{+1}$ of the level $^3S_1$ to level $^3P_0$. At $\theta=0^{\circ}$, the only probed transition is $\ket{m_x=0}\leftrightarrow \textcolor{black}{\ket{m_x=0}}$, which leads to a strong optical pumping of the atoms to the $\ket{m_x=\pm1}$ sublevels and an increased transmission. For $\theta=55^{\circ}$ the three transitions see equal Rabi frequencies, so that the absorption is maximum.

This polarization dependent shot noise is taken into account in the simulations using the absorption measurements of Fig. \ref{ShotNoise}(a). Taking into account the $S=0.7$\,A/W sensitivity of our balanced detection, its $0.5\times10^4$\,V/A  transimpedance gain, and a 9\,MHz bandwidth, leads to the noise simulations in  Figs. \ref{ShotNoise}(c) and \ref{ShotNoise}(d) performed over a time sequence of 1\,s with $18\times 10^6$ points.  The $\theta$ dependence of the shot noise is hardly visible in the first case (1.5\,GHz detuning), but becomes very clear in the second (0.3\,GHz detuning). These shot noise levels are consistent with the experimental ones.

\begin{figure}[h]
    \centering
    \includegraphics[width=\columnwidth]{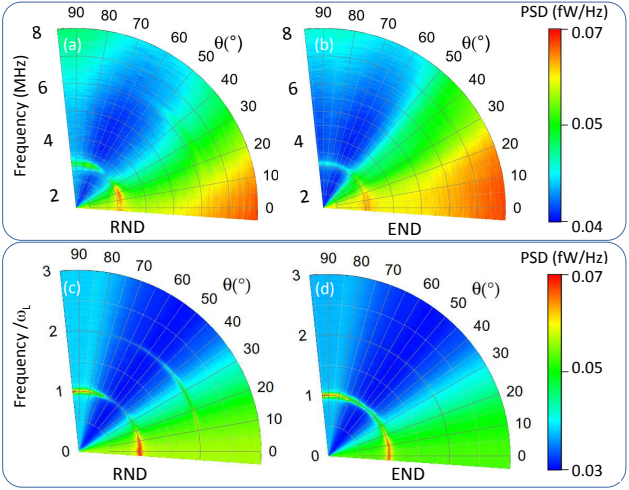}
    \caption{(a,b) Experimental and (c,d) simulated spin noise spectra obtained 300\,MHz above the $D_0$ transition for a 1\,mW laser input power and a 30\,MHz Rabi frequency respectively. (a) and (c) correspond to the RND set-up while (b) and (d) are obtained with the END one. The other parameters are the same as in Fig.\ref{fig.PolarizationDependence} .} 
    \label{pienear}
\end{figure}

Repeating measurements similar to those of Figs.\,\ref{3deg}(c) and \ref{3deg}(d) for varying values of $\theta$ leads to Figs.\,\ref{pienear}(a) and (b) for RND and END noises, respectively, for an optical detuning of 300\,MHz. The corresponding simulation results, including the calculated polarization dependent shot noise, are shown in Fig. \ref{pienear}(c) and (d). A very good agreement is obtained between theory and experiments.

In spite of the polarization dependent noise floor discussed above, the results of Fig.\,\ref{pienear} clearly show that the polarization dependence of the RND and END spectra have reversed their behaviors close to resonance (see Fig.\,\ref{pienear}) with respect to the far detuned case (see Fig.\,\ref{fig.PolarizationDependence}). Indeed, the  RND noise peaks of Fig.\,\ref{pienear} now exhibit a strong $\theta$ dependence, while the END noise peak amplitude becomes almost independent of the polarization. Moreover, the peak at $2\omega_L$ has disappeared from the END spectra, while a weak one indeed appears a bit above 6\,MHz in the RND spectra for angles $\theta$ below $50^{\circ}$, as also evidenced in  Fig.\,\ref{3deg}(c).  

This exchange between the END and RND behaviors at large and small optical detunings is well explained by the fact that dispersive effects dominate at large detunings while absorption effects dominate close to resonance.  Far from resonance, circular and linear birefringence effects are dominant, leading to polarization independent results for the RND set-up and polarization dependent spectra for the END one respectively. On the contrary, close to resonance, circular dichroism leads to polarization independent END spectra (except for the shot noise) while linear dichroism is associated with $\theta$ dependent RND spectra. Petrov \& al \cite{Petrov18a} demonstrated that due to slow spin relaxation dynamics, the spin noise signal in cesium atoms vanishes at the center of the Doppler broadened transition, that behaves like a homogeneous one. Helium atoms behaviour is similar for conventional birefringence noise, and we indeed model the Doppler broadening with a simple decay rate in our numerical model. But due to the simple level structure of helium, which avoids to mix effects from different lines, we can detect dichroism noise close to resonance.

\section{Interpretation in terms of spin oscillation modes}\label{Sec06}
Some physical insight can be gained in the existence of the dual-peak noise spectrum if one remembers that the probed ground state is a spin\,1 system. Let us indeed determine the oscillation modes of the density matrix of the system restricted to its three ground state sublevels submitted to a magnetic field, without any decay nor noise, for different initial states. With the quantization axis oriented along $z$,  the evolution of this system  under the influence of the magnetic field oriented along $x$ is governed by the Hamiltonian of Eq.\,(\ref{Eq07}) restricted to its first three lines and columns with $\Omega^+=\Omega^-=0$.

\begin{figure*}[]
    \centering
    \includegraphics[width=0.8\textwidth]{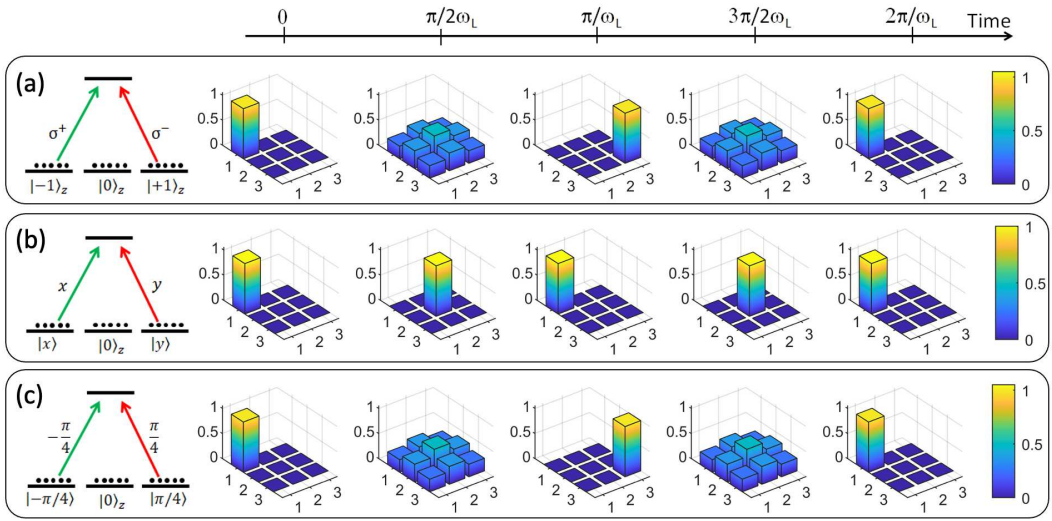}
    \caption{Free evolution of the moduli of the density matrix of the three ground sublevels in the presence of a transverse magnetic field but without any light, for 3 different initial conditions: (a) all the atoms are initially in the state $\ket{-1}_{z}$, which can be coupled to the excited state by circularly polarized light; (b) all the atoms are initially in the state $\ket{x}$, which can be coupled to the excited state by polarized light along the B-field orientation $x$; (c) all the atoms are initially in the state $\ket{-\pi/4}$, which can be coupled to the excited state by linearly polarized light at $\pi/4$ from the B-field orientation $x$. In the cases (a) and (c), the populations oscillate at the frequency $\omega_L$, while the oscillation frequency is $2\omega_L$ in the case (b).}
    \label{fig.OscillationModes}    
\end{figure*}
Figures \ref{fig.OscillationModes}(a), (b) and (c) show the time evolution of the system density matrix for three different initial conditions, over a duration $\pi/\omega_{\mathrm{L}}$ and plotted in three different basis for the ground state sublevels. In the case of Fig.\,\ref{fig.OscillationModes}(a) the system is initially in the state $\left|-1\right\rangle_z$. 
In the case of Fig.\,\ref{fig.OscillationModes}(b) the system is initially in the superposition state $\left|x\right\rangle=(\left|-1\right\rangle_z+\left|1\right\rangle)_z/\sqrt{2}$. Finally, in the case of Fig.\,\ref{fig.OscillationModes}(c) the system is initially in  $\left|-\pi/4\right\rangle=(e^{i\pi/4}\left|-1\right\rangle_z+e^{-i\pi/4}\left|+1\right\rangle_z)/\sqrt{2}$. In the first and last cases, the system evolves periodically at  frequency $\omega_{\mathrm{L}}$, while it evolves at frequency $2\omega_{\mathrm{L}}$ in the second one.

Each of these three different oscillation modes of the spin-1 system can be observed or not depending on the probe light polarization. For example, the oscillation mode of Fig.\,\ref{fig.OscillationModes}(a) corresponds to an oscillation of the system between the states $\left|-1\right\rangle_z$ and $\left|+1\right\rangle_z$. This means that the system exhibits a magnetization along $z$ oscillating at frequency $\omega_{\mathrm{L}}$: when it interacts with far detuned linearly polarized light, the $\sigma^+$ and $\sigma^-$ components of light see an oscillating refractive index difference, that corresponds to a Faraday rotation oscillating at $\omega_{\mathrm{L}}$, which is present whatever the orientation $\theta$ of the incident polarization, as observed in  Figs.\,\ref{fig.PolarizationDependence}(a) and \ref{fig.PolarizationDependence}(c).

The second mode of oscillation (see Fig.\,\ref{fig.OscillationModes}(b)) is plotted in the basis formed by states $\left|x\right\rangle$, $\left|y\right\rangle=i(\left|-1\right\rangle_z-\left|+1\right\rangle)_z/\sqrt{2}$, and $\left|0\right\rangle_z$.  It corresponds to an oscillation of the system between the state $\left|x\right\rangle$ and the state $\left|0\right\rangle_z$ at frequency $2\omega_{\mathrm{L}}$. Since  $\left|x\right\rangle$ (resp. $\left|y\right\rangle$) is the state that maximizes the coupling to $x-$ (resp $y-$) polarized light, this mechanism thus creates a linear birefringence oscillating at frequency $2\omega_{\mathrm{L}}$ with neutral axes oriented along $x$ and $y$ for far detuned light. The ellipticity imprinted on an incident linearly polarized light is thus maximum for $\theta=45^{\circ}$: this explains why the peak at frequency $2\omega_{\mathrm{L}}$ in Figs.\,\ref{fig.PolarizationDependence}(b) and \ref{fig.PolarizationDependence}(d) is detected only as ellipticity noise when the probe polarisation direction is around $45^{\circ}$ with respect to $x$ and $y$.

Similarly, the oscillation mode of Fig.\,\ref{fig.OscillationModes}(c) corresponds to an oscillation of the spin state between states $\left|\pi/4\right\rangle=(e^{-i\pi/4}\left|-1\right\rangle_z+e^{i\pi/4}\left|+1\right\rangle_z)/\sqrt{2}$ and $\left|-\pi/4\right\rangle$ at frequency $\omega_{\mathrm{L}}$. These states are those that maximize the excitation by linear polarization oriented at $+\pi/4$ and $-\pi/4$ with respect to $x$, respectively. This oscillating modes thus creates an oscillating birefringence with neutral axes at $\pm \pi/4$ with respect to $x$ and $y$. One thus understands why ellipticity noise at frequency $\omega_{\mathrm{L}}$ can be observed in Figs.\,\ref{fig.PolarizationDependence}(b) and \ref{fig.PolarizationDependence}(d) only when the incident polarisation is close to $\theta=0$ and $\theta=\pi/2$ and not around  $\pm \pi/4$.

Of course, as discussed in Sec.\,\ref{Sec05}, when one gets close to resonance, birefringence (dispersive effects) become negligible compared to dichroism (absorption) effects, explaining the change to the behavior of Figs.\, \ref{3deg} and \ref{pienear}.

\section{Conclusion}
We have investigated the spin noise spectra of a thermal vapor of metastable $^4$He atoms at room temperature using a probe beam whose frequency is tuned close to the $D_{0}$ line. Indeed, this line offers the advantage of being a very pure spin 1 system, which had not yet been explored in the context of spin noise studies. Using two different setups, we could record the signature of spin fluctuations giving rise to polarization rotation or ellipticity fluctuations. Taking advantage of the simplicity of the level structure of metastable helium, we could choose to operate far from or close to resonance, without disturbance from neighboring transitions. In the first case, the well-known and polarization independent SNS resonance was observed at the Larmor frequency $\omega_{\mathrm{L}}$ in the polarization rotation noise spectra, while we could confirm that the ellipticity noise spectra exhibit two polarization dependent resonance peaks at the Larmor and at twice the Larmor frequencies \cite{Bretenaker2019,Fomin20}. In the second case, near resonance, we have shown that the behaviors are exchanged: ellipticity noise spectra exhibit only one polarization independent peak while the rotation noise spectra have two peaks whose magnitudes depend on the input polarization orientation. These results are interpreted as the result of a competition between linear and circular dichroism and birefringence effects.

These observations were successfully confirmed by numerical simulations of the system based on a simple model in which spin noise is created by the transit of the atoms through the laser beam. Both qualitative and even quantitative  agreement is obtained with the experimental results. It should be noted that although the parameters are extracted from other experimental data, the computed noise values agree with the recorded spin noise spectra. We also gave an interpretation of the spin noise peaks and their polarization behavior based on the existence of three spin oscillation modes that can be probed or not depending on the input polarization orientation.

Future theoretical work might better take into account the effect of the Doppler broadening and the fact that the medium becomes optically thick when the probe is close to resonance. Various fluctuation sources should also be considered: the well separated transitions of metastable helium can again be an asset for a deep understanding of the contributions of the effects affecting the spin noise spectra of a spin 1.

\selectlanguage{english}

\begin{acknowledgments}
The authors acknowledge funding by the Institut Universitaire de France and the Labex PALM, and thank D. Roy, S. Chaudhury, and H. Ramachandran for fruitful discussions.
\end{acknowledgments}

\bibliography{bibliography}
\bibliographystyle{apsrev4-1}

\appendix

\end{document}